# Mathematical Inconsistencies in Dirac Field Theory

by

Dan Solomon
Rauland-Borg Corporation
3450 W. Oakton
Skokie, IL 60076

Please send all correspondence to:

Dan Solomon
1604 Brummel
Evanston, IL USA 60202

Phone: 847-679-0900 ext. 5337
Email: dsolomon@northshore.net

## Abstract

If a mathematical theory contains incompatible postulates then it is likely that the theory will produce theorems or results that are contradictory. It will be shown that this is the case with Dirac field theory. An example of such a contradiction is the problem associated with the evaluating the Schwinger term $\langle 0|[\hat{\rho}(\vec{x}),\vec{J}(\vec{y})]|0\rangle$. It is generally known that different ways of evaluating this quantity yield different results. It will be shown that the reason for this is that Dirac field theory is mathematically inconsistent, i.e. , it contains incompatible assumptions or postulates. The generally accepted definition of the vacuum state must be modified in order to create a consistent theory.

## 1. Introduction

The starting point for a mathematical theory is a set of postulates. These are mathematical statements which are assumed to be true without proof. From these postulates additional mathematical statements called theorems can be formulated. One question that arises in specifying a set of postulates is whether or not the postulates are internally consistent. This is often not readily apparent. If the postulates are not consistent then they will lead to contradictions and the theory will contain theorems that contradict each other.

In this article it will be shown that this is indeed the situation that exists in Dirac field theory. That is, the theory is mathematically inconsistent in that it contains basic postulates that lead to contradictory results. One example of this is the calculation of the equal time commutator $\left[\hat{\rho}(\vec{x}),\vec{J}(\vec{y})\right]$ where $\hat{\rho}$ is the charge operator and $\hat{\vec{J}}$ is the current operator. It can be shown that when this commutator is evaluated using the equal time anticommutator equations for the field operators then the result is zero (see appendix 3 of Heitler [1]). However when the quantity $\left\langle 0\left|\left[\hat{\rho}(\vec{x}),\vec{J}(\vec{y})\right]\right|0\right\rangle$ is calculated, using the definitions of the field operators and their action on the vacuum state $\left|0\right\rangle$, it is known that the result is not zero. This non-zero result is called the Schwinger term (see Pauli and Villiars [2], Schwinger[3], Boulware and Jackiw[4], Page 449 of Weinberg[5]). Thus two different ways of calculating a given quantity produce contradictory results.

One way that this inconsistency manifests itself is in the calculation of the vacuum current. Dirac field theory is assumed to be gauge invariant. However it is well known that a direct calculation of the first order change in the vacuum current due to an applied external field does not produce a gauge invariant result. (See, for example, Chapt. 14 of Greiner et al[6], Chap. 2 of Pauli[7], Section 32 of Heitler [1], and Chap. 5 of Greiner and Reinhardt[8]). To make the theory gauge invariant the non-gauge invariant terms must be identified and removed from the expression for the vacuum current. This can be done simply by examining the expression for the vacuum current and removing "by hand" the terms with obvious non-gauge invariant and non-physical characteristics. A formal method of dealing with this problem is the method of invariant regulation of Pauli and Villiars [2]. In this method other functions are introduced that happen to have the correct behavior so that the non-gauge invariant terms are cancelled. However, as Pauli and Villars point out, this is merely a mathematical device and there is no physical justification for this method.

The question that arises is "where is the gauge invariance lost?" That is, we start out with a theory that is assumed to be gauge invariant. However, when we use the theory to calculate the vacuum current in the presence of an applied electric field we get a non-gauge invariant result. There does not appear to be a good explanation in the literature as to why this problem occurs except for a commonly held belief that it is due to some artifact of the mathematics (see comments on page 398 of Greiner et al [6]). However new light was shed on this problem in a recent paper by this author (see Solomon [9]). It was shown that for Dirac theory to be gauge invariant puts a certain requirement on the vacuum state. The requirement is that the vacuum state cannot be the minimum energy state. The vacuum state must be defined to allow for the existence of basis states with less energy than the vacuum. If this condition is not satisfied then the gauge invariance of the theory is destroyed. The normal vacuum state vector $|0\rangle$ does not meet this requirement. The way to formulate a vacuum state that will meet the requirement for gauge invariance was discussed in Ref [9] and is briefly reviewed here, as follows.

The solution of the single particle Dirac equation includes both positive and negative energy states. The vacuum state $|0\rangle$ is normally defined as the quantum state in which all the negative energy states are occupied by a single electron. In Ref [9] it was shown that $|0\rangle$ does not lead to a gauge invariant theory. In order to achieve gauge invariance the definition of the vacuum state was modified as follows. Consider the quantum state where all negative energy states from the top of the negative energy band, at an energy of $-m$ (where m is the mass), to an energy of $-(m+\Delta E_w)$ are occupied by a single electron. All positive energy states, and all negative energy states with energy less than $-(m+\Delta E_w)$ are unoccupied. Represent this quantum state by $|0,\Delta E_w\rangle$. The modified vacuum state is defined as the state $|0,\Delta E_w\rangle$ where $\Delta E_w \to \infty$. At first glance this seems almost the same as the original vacuum state $|0\rangle$. That is, all negative energy states from $-m$ to $-\infty$ are occupied by a single electron. The critical difference is that when calculations are done using the state $|0,\Delta E_w \to \infty\rangle$ the parameter $\Delta E_w$ is treated as a finite number during the calculation and only set to infinity at the end of the calculation. When this is done, as is shown in Ref [9], the Schwinger term $\langle 0,\Delta E_w \to \infty \left| \left[ \hat{\rho}(\vec{x}), \hat{\vec{J}}(\vec{y}) \right] \right| 0,\Delta E_w \to \infty \rangle$ will be zero and the expression for the vacuum current will be gauge invariant.

In this article we will extend the results of Ref [9] . In particular it will be shown that the problems that are associated with the evaluation of $\langle 0 \| [\hat{\rho}(\vec{x}), \hat{\vec{J}}(\vec{y})] \| 0 \rangle$ are due to the fact that, in its present formulation, Dirac field theory is mathematically inconsistent. By this I mean that there exist within the theory incompatible assumptions or postulates. The result is that there are different methods of computing $\langle 0 \| [\hat{\rho}(\vec{x}), \hat{\vec{J}}(\vec{y})] \| 0 \rangle$ which give different results.

## 2. An Inconsistency

There is a well known inconsistency associated with the commutator $[\hat{\rho}(\vec{x}), \hat{\vec{J}}(\vec{y})]$ which as been discussed by previous authors (see Schwinger[3], Johnson[10], and page 530 of Itzykson and Zuber[11]) and will be examined in this section. In the following discussion it will be convenient to work in the Schrodinger representation. In this case the field operators $\hat{\psi}(\vec{x})$ are time independent and satisfy the anticommutator relationship

$$\hat{\psi}_a^\dagger(\vec{x})\hat{\psi}_b(\vec{y}) + \hat{\psi}_b(\vec{y})\hat{\psi}_a^\dagger(\vec{x}) = \delta_{ab}\delta^{(3)}(\vec{x}-\vec{y}) \tag{1}$$

All other anticommutators =0

$\hat{\vec{J}}$ and $\hat{\rho}$ are given in terms of the field operator $\hat{\psi}(\vec{x})$ according to

$$\hat{\vec{J}}(\vec{x}) = \frac{e}{2}\left[\hat{\psi}^\dagger(\vec{x}), \vec{\alpha}\hat{\psi}(\vec{x})\right] \tag{2}$$

$$\hat{\rho}(\vec{x}) = \frac{e}{2}\left[\hat{\psi}^\dagger(\vec{x}), \hat{\psi}(\vec{x})\right] \tag{3}$$

where ‘e’ is the electric charge. Also define the free field Hamiltonian operator

$$\hat{H}_o = \frac{1}{2}\int\left[\hat{\psi}^{\dagger}(\vec{x}), H_o\hat{\psi}(\vec{x})\right]d\vec{x} - \xi_R \tag{4}$$

In this expression $\xi_R$ is a renormalization constant which can be defined so that the energy of the vacuum state is zero and

$$H_o = -i\vec{\alpha}\cdot\vec{\nabla} + m\beta \tag{5}$$

where m is the mass and $\vec{\alpha}$ and $\beta$ are the usual 4x4 matrices.

A straight forward application of (1) yields the following relationships

$$\left[\hat{\rho}(\vec{y}), \hat{\vec{J}}(\vec{x})\right] = 0 \tag{6}$$

$$\left[\hat{\rho}(\vec{y}), \hat{\rho}(\vec{x})\right] = 0 \tag{7}$$

and

$$\left[\hat{H}_o, \hat{\rho}(\vec{x})\right] = i\vec{\nabla}\cdot\hat{\vec{J}}(\vec{x}) \tag{8}$$

These are well know results that follow from (1) or symmetry arguments. (See appendix 3 of Heitler[1] for (6) and (7) and see Johnson[10] or page 530 of Ref[11] for (8) which is the continuity equation). The proof that these equations follow from (1) is given here in Appendix A.

It will be convenient to define the weighted charge operator by

$$\hat{\rho}_w \equiv \int \hat{\rho}(\vec{x})\chi(\vec{x})d\vec{x} \tag{9}$$

where $\chi(\vec{x})$ is an arbitrary real valued function.

Use this in (6) and (8) to obtain

$$\left[\hat{\rho}_w, \hat{\vec{J}}(\vec{x})\right] = 0 \tag{10}$$

and

$$\left[\hat{H}_o, \hat{\rho}_w\right] = i\int \chi(\vec{x})\nabla \cdot \hat{\vec{J}}(\vec{x})d\vec{x} \tag{11}$$

Assume reasonable boundary conditions at $|\vec{x}| \to \infty$ and integrate by parts to obtain

$$\left[\hat{H}_o, \hat{\rho}_w\right] = -i\int \hat{\vec{J}}(\vec{x}) \cdot \vec{\nabla}\chi(\vec{x})d\vec{x} \tag{12}$$

Use this result to obtain the expression

$$\left[\hat{\rho}_w, \left[\hat{H}_o, \hat{\rho}_w\right]\right] = -i\left[\hat{\rho}_w, \int \hat{\vec{J}}(\vec{x}) \cdot \vec{\nabla}\chi(\vec{x})d\vec{x}\right] \tag{13}$$

Rearrange terms to obtain

$$\left[\hat{\rho}_w, \left[\hat{H}_o, \hat{\rho}_w\right]\right] = -i\int \left[\hat{\rho}_w, \hat{\vec{J}}(\vec{x})\right] \cdot \vec{\nabla}\chi(\vec{x})d\vec{x} \tag{14}$$

Next use (10) in the above to yield

$$\left[\hat{\rho}_w, \left[\hat{H}_o, \hat{\rho}_w\right]\right] = 0 \tag{15}$$

Now let $|\varphi_n\rangle$ be the set of eigenvectors of the operator $\hat{H}_o$ with energy eigenvalue of $\varepsilon_n$. The $|\varphi_n\rangle$ form a set of basis states in a fock space (see Chapt. 3 of Itzykson and Zuber[11] ) and satisfy the equations

$$\hat{H}_o|\varphi_n\rangle = \varepsilon_n|\varphi_n\rangle; \quad \langle\varphi_n|\hat{H}_o = \langle\varphi_n|\varepsilon_n \tag{16}$$

and

$$\langle\varphi_n|\varphi_m\rangle = \delta_{nm} \tag{17}$$

Also we can define the relationship

$$\sum_n |\varphi_n\rangle\langle\varphi_n| = 1 \tag{18}$$

(see Chapter VII of Messiah [12]).

The vacuum state $|0\rangle$ is generally assumed to be the eigenvector of $\hat{H}_o$ with the smallest eigenvalue. Let $\varepsilon_o$ be the eigenvalue of $|0\rangle$. Adjust the renormalization constant $\xi_R$ so that $\varepsilon_o = 0$, then we can write

$$\varepsilon_n > \varepsilon_o = 0 \text{ for } |\varphi_n\rangle \neq |0\rangle \tag{19}$$

Next we will consider two ways to evaluate the expression $\frac{1}{2}\langle 0|\left[\hat{\rho}_w,\left[\hat{H}_o,\hat{\rho}_w\right]\right]|0\rangle$. First use (15) to obtain

$$\frac{1}{2}\langle 0|\left[\hat{\rho}_w,\left[\hat{H}_o,\hat{\rho}_w\right]\right]|0\rangle = 0 \tag{20}$$

The above result follows directly from the anticommutator relationships given in (1).

Now we will consider another way to evaluate $\frac{1}{2}\langle 0|\left[\hat{\rho}_w,\left[\hat{H}_o,\hat{\rho}_w\right]\right]|0\rangle$ which will yield a different result. Use (16) to obtain

$$\begin{aligned}\frac{1}{2}\langle 0|\left[\hat{\rho}_w,\left[\hat{H}_o,\hat{\rho}_w\right]\right]|0\rangle &= \frac{1}{2}\langle 0|\left(\hat{\rho}_w\left(\hat{H}_o\hat{\rho}_w - \hat{\rho}_w\hat{H}_o\right) - \left(\hat{H}_o\hat{\rho}_w - \hat{\rho}_w\hat{H}_o\right)\hat{\rho}_w\right)|0\rangle \\ &= \langle 0|\hat{\rho}_w\hat{H}_o\hat{\rho}_w|0\rangle - \varepsilon_o\langle 0|\hat{\rho}_w\hat{\rho}_w|0\rangle\end{aligned} \tag{21}$$

Next use the fact that $\varepsilon_o = 0$ to obtain

$$\frac{1}{2}\langle 0|\left[\hat{\rho}_w,\left[\hat{H}_o,\hat{\rho}_w\right]\right]|0\rangle = \langle 0|\hat{\rho}_w\hat{H}_o\hat{\rho}_w|0\rangle \tag{22}$$

Use (16) through (18) to yield

$$\langle 0|\hat{\rho}_w\hat{H}_o\hat{\rho}_w|0\rangle = \sum_{n,m}\langle 0|\hat{\rho}_w|\varphi_n\rangle\langle\varphi_n|\hat{H}_o|\varphi_m\rangle\langle\varphi_m|\hat{\rho}_w|0\rangle = \sum_n\left|\langle 0|\hat{\rho}_w|\varphi_n\rangle\right|^2\varepsilon_n \tag{23}$$

Now, in the above equation the quantity $\langle 0|\hat{\rho}_{w}|\varphi_{n}\rangle$ is in general non-zero and according to (19) $\varepsilon_{n}$ is non-negative. Therefore the quantity on the right side of the above equation is greater than zero. Use this result in (22) to obtain

$$\frac{1}{2}\langle 0|\left[\hat{\rho}_{w},\left[\hat{H}_{o},\hat{\rho}_{w}\right]\right]|0\rangle > 0 \tag{24}$$

This result is direct conflict with (20). Thus there is an inconsistency.

The generally accepted explanation as to why two different methods of calculation give different results for the same quantity is that the behavior of the quantity $\langle 0|\left[\hat{\rho}(\vec{y}),\hat{\vec{J}}(\vec{x})\right]|0\rangle$ is so singular that a "naive" calculation using field operator anticommutator relations cannot be trusted and that the commutator $\left[\hat{\rho}(\vec{y}),\hat{\vec{J}}(\vec{x})\right]$ is actually nonzero. The problem with this idea is that $\left[\hat{\rho}(\vec{y}),\hat{\vec{J}}(\vec{x})\right]$ must be zero for Dirac field theory to be gauge invariant and have local charge conservation. This will be demonstrated in Sections 4 and 5.

A careful examination of the discussion leading up to (20) and (24) suggests another explanation for this problem. Equation (20) is derived from (10) and (12) which in turn are ultimately derived from the field operator anticommutator relations (Equation (1)). Equation (20), then, follows directly from (1). If (1) is a considered to be a postulate then (20) is a theorem derived from this postulate. In deriving (24), on the other hand, (1) is not used. Equation (24) follows from the properties of the eigenvectors of $\hat{H}_{o}$ as defined in (16) through (19). The key factor in deriving (24) is (19), i.e., the assumption that energy eigenvalue of the basis states $|\varphi_{n}\rangle$ is always positive (if the energy of the vacuum state $|0\rangle$ is set to zero). If (19) is considered to be a postulate then (24) is a theorem derived from this postulate. The fact that that (20) and (24) are contradictory suggests that the underlying postulates (1) and (19) are not compatible. It will be shown that this is, indeed, the case and that this inconsistency can be removed by modifying the definition of the vacuum state.

In order to resolve this contradiction the vacuum state must be defined in such a way so that (23) is zero, i.e,

$$\langle 0|\hat{\rho}_{w}\hat{H}_{o}\hat{\rho}_{w}|0\rangle = \sum_{n}\left|\langle 0|\hat{\rho}_{w}|\varphi_{n}\rangle\right|^{2}\varepsilon_{n} = 0 \tag{25}$$

Clearly in order for this equation to be true our theory must allow for the existence of basis states whose energy eigenvalue $\varepsilon_n$ is negative with respect to the vacuum state. It was shown in ref. [9] that this can be achieved by using the state vector $|0,\Delta E_{w}\rightarrow\infty\rangle$ as the vacuum state. In order that this paper can be self contained the following section on the vacuum state is similar to the corresponding discussion in ref. [9].

## 3. Redefining the vacuum state

At this point we shall briefly review the definition of the vacuum state $|0\rangle$ as it is normally defined in Dirac field theory. The field operator can be expressed according to

$$\hat{\psi}(\vec{x}) = \sum_{n}\hat{a}_{n}\phi_{n}(\vec{x}); \quad \hat{\psi}^{\dagger}(\vec{x}) = \sum_{n}\hat{a}_{n}^{\dagger}\phi_{n}^{\dagger}(\vec{x}) \tag{26}$$

where the $\hat{a}_n$ ($\hat{a}_n^{\dagger}$) are the destruction(creation) operators for a particle in the state $\phi_n(\vec{x})$. The operators $\hat{a}_n$ and $\hat{a}_n^{\dagger}$ along with the function $\phi_n(\vec{x})$ are defined in such a way so that (26) is consistent with (1). The operators $\hat{a}_n$ and $\hat{a}_n^\dagger$ satisfy the anticommutator equation

$$\begin{aligned}&\hat{a}_n^\dagger \hat{a}_m + \hat{a}_m \hat{a}_n^\dagger = \delta_{mn} \\ &\text{All other anticommutators} = 0\end{aligned} \tag{27}$$

The $\phi_n(\vec{x})$ are eigenfunctions of the free field single particle Dirac equation with energy eigenfunction $\lambda_n E_n$. That is,

$$\lambda_n E_n \phi_n(\vec{x}) = H_o \phi_n(\vec{x}) \tag{28}$$

where

$$E_n = +\sqrt{\vec{p}_n^2 + m^2}, \quad \lambda_n = \left\langle \begin{array}{l} +1 \text{ for a positive energy state} \\ -1 \text{ for a negative energy state} \end{array} \right. \tag{29}$$

and where $\vec{p}_n$ is the momentum of the state n. Solutions of (28) are of the form

$$\phi_n(\vec{x}) = u_n e^{i\vec{p}_n \cdot \vec{x}} \tag{30}$$

where $u_n$ is a constant 4-spinor. The $\phi_n(\vec{x})$ form a complete orthonormal basis in Hilbert space and satisfy

$$\sum_n \left(\phi_n^\dagger(\vec{x})\right)_{(a)} \left(\phi_n(\vec{y})\right)_{(b)} = \delta_{ab} \delta^3(\vec{x} - \vec{y}) \tag{31}$$

where “a” and “b” are spinor indices (see page 107 of Heitler [1]).

Following chapter 9 of Greiner et al[6] define the state vector $|0, \text{bare}\rangle$ which is the state vector that is empty of all particles, i.e.,

$$\hat{a}_n |0, \text{bare}\rangle = 0 \text{ for all n} \tag{32}$$

The operator $\hat{a}_n$ ($\hat{a}_n^\dagger$) destroys(creates) a particle with energy $\lambda_n E_n$. For the index ‘n’ we will define the following

n<0 refers to negative energy states. (33)

n>0 refers to positive energy states.

The vacuum state vector $|0\rangle$ is defined as the state vector in which all negative energy states are occupied by a single particle. Therefore

$$|0\rangle = \prod_{n<0} \hat{a}_n^\dagger |0, \text{bare}\rangle \tag{34}$$

where, as defined above, n<0 means that the product is taken over all negative energy states. From this expression, and (27) and (32), $|0\rangle$ can then be defined by

$$\hat{a}_n|0\rangle = 0 \text{ for } n > 0\ ;\ \hat{a}_n^\dagger|0\rangle = 0 \text{ for } n < 0 \tag{35}$$

As has been discussed in the last section a mathematically consistent theory requires that there exist negative energy basis states. To see how this can be achieved recall the definition of the state vector $|0, \Delta E_w\rangle$ as described in the introduction. $|0, \Delta E_w\rangle$ is the state vector for which each single particle state in the band of negative energy states from energy $-m$ to $-(m+\Delta E_w)$ is occupied and all other single particle states are unoccupied. Let the notation "n∈band" mean that "n" is a single particle quantum state with energy in the range $-m$ to $-(m+\Delta E_w)$. The notation "n<band" means that "n" is a single particle quantum state with energy less than $-(m+\Delta E_w)$. Recall, also, that n>0 refers to positive energy states. Therefore $|0, \Delta E_w\rangle$ can be defined by

$$\begin{aligned} &\hat{a}_n|0, \Delta E_w\rangle = 0 \text{ for } n > 0 \\ &\hat{a}_n^\dagger|0, \Delta E_w\rangle = 0 \text{ for } n \in \text{band} \\ &\hat{a}_n|0, \Delta E_w\rangle = 0 \text{ for } n < \text{band} \end{aligned} \tag{36}$$

$\left|0,\Delta E_w\right\rangle$ is an eigenvector of the operator $\hat{H}_o$. The eigenvalue can be set equal to zero by proper selection of the renormalization constant $\xi_R$. Therefore we write,

$$\hat{H}_o\left|0,\Delta E_w\right\rangle = 0 \tag{37}$$

(Note that defining the energy as zero is a mathematical convenience that merely specifies an energy reference point. The results would not change if the energy reference point was changed.)

Now let $m \in$ band and $n$<band. The operator pair $\hat{a}_n^{\dagger}\hat{a}_m$ acting on $\left|0,\Delta E_w\right\rangle$ will destroy a particle with the quantum number 'm' within the band and produce a particle with quantum number 'n' underneath the band. This will produce a quantum state with energy less then $\left|0,\Delta E_w\right\rangle$ by the amount $\left(E_n - E_m\right)$. Therefore, if $m \in$ band and $n$<band, then $\hat{a}_n^{\dagger}\hat{a}_m\left|0,\Delta E_w\right\rangle$ is an example of a negative energy basis state. For the state $\left|0\right\rangle$ such transitions are not allowed. The action of the operators $\hat{a}_n^{\dagger}$ or $\hat{a}_m$ on $\left|0\right\rangle$ is to either destroy $\left|0\right\rangle$, or create a previously unoccupied positive energy state or destroy an occupied negative energy state. In the latter two cases the energy of the state vector $\left|0\right\rangle$ is increased.

Refer back to the discussion in Section 2 and replace $|0\rangle$ by $\left|0,\Delta E_w\right\rangle$. Recall that we require that the quantity $\left\langle 0,\Delta E_w\right|\rho_w\hat{H}_o\rho_w\left|0,\Delta E_w\right\rangle$ to be equal to zero if the theory is to be mathematically consistent (see (25)). This will be proven as follows

Use (3), (9), (26), and (27) to obtain

$$\hat{\rho}_W = \frac{e}{2}\left\{\sum_{m,n}\left(\int \phi_n^\dagger \chi \phi_m d\vec{x}\right)\left(\hat{a}_n^\dagger \hat{a}_m - \hat{a}_m \hat{a}_n^\dagger\right)\right\} = e\left\{\sum_{m,n}\left(\int \phi_n^\dagger \chi \phi_m d\vec{x}\right)\left(\hat{a}_n^\dagger \hat{a}_m - \frac{\delta_{mn}}{2}\right)\right\} \tag{38}$$

Therefore,

$$\begin{aligned}\hat{\rho}_W\left|0,\Delta E_W\right\rangle &= \sum_{\substack{m\in band\\ n>0}} e\left(\int \phi_n^\dagger \chi \phi_m d\vec{x}\right)\hat{a}_n^\dagger \hat{a}_m\left|0,\Delta E_W\right\rangle \\ &+ \sum_{\substack{m\in band\\ n<band}} e\left(\int \phi_n^\dagger \chi \phi_m d\vec{x}\right)\hat{a}_n^\dagger \hat{a}_m\left|0,\Delta E_W\right\rangle + \rho'_{W,vac}\left|0,\Delta E_W\right\rangle\end{aligned} \tag{39}$$

where $\rho'_{v,wac}$ is a constant given by

$$\rho'_{v,wac} = \sum_{m\in band} e\left(\int \phi_m^\dagger \chi \phi_m d\vec{x}\right) - \frac{e}{2}\sum_{all\ m}\left(\int \phi_m^\dagger \chi \phi_m d\vec{x}\right) = \left\langle 0,\Delta E_W\right|\hat{\rho}_W\left|0,\Delta E_W\right\rangle \tag{40}$$

The action of the operator $\hat{a}_m$ destroys a particle with energy $\lambda_m E_m$ and the action the operator $\hat{a}_n^\dagger$ creates a particle with energy $\lambda_n E_n$. Therefore

$$\hat{H}_o \hat{a}_n^\dagger \hat{a}_m\left|0,\Delta E_W\right\rangle = \left(\lambda_n E_n - \lambda_m E_m\right)\hat{a}_n^\dagger \hat{a}_m\left|0,\Delta E_W\right\rangle \tag{41}$$

Use this result along with (37) and the hermitian conjugate of (39) to obtain

$$\begin{aligned}\left\langle 0,\Delta E_W\right|\hat{\rho}_W \hat{H}_o \hat{\rho}_W\left|0,\Delta E_W\right\rangle &= \sum_{\substack{m\in band\\ n>0}} e^2\left|\left(\int \phi_n^\dagger \chi \phi_m d\vec{x}\right)\right|^2\left(E_m + E_n\right) \\ &+ \sum_{\substack{m\in band\\ n<band}} e^2\left|\left(\int \phi_n^\dagger \chi \phi_m d\vec{x}\right)\right|^2\left(E_m - E_n\right)\end{aligned} \tag{42}$$

Next define

$$F_1 \equiv \sum_{\substack{m\in band \\ n\in band}} e^2 \left| \left( \int \phi_n^\dagger \chi \phi_m d\vec{x} \right) \right|^2 (E_m - E_n) \tag{43}$$

The dummy indices 'n' and 'm' can be switched in the above to obtain

$$F_1 = \sum_{\substack{m\in band \\ n\in band}} e^2 \left| \left( \int \phi_m^\dagger \chi \phi_n d\vec{x} \right) \right|^2 (E_n - E_m) \tag{44}$$

Next use the fact that

$$\left| \left( \int \phi_m^\dagger \chi \phi_n d\vec{x} \right) \right|^2 = \left( \int \phi_m^\dagger \chi \phi_n d\vec{x} \right)\left( \int \phi_n^\dagger \chi \phi_m d\vec{x} \right) = \left| \left( \int \phi_n^\dagger \chi \phi_m d\vec{x} \right) \right|^2 \tag{45}$$

in the above to yield

$$F_1 = -F_1 = 0 \tag{46}$$

Since $F_1=0$ it can be added to (42) to obtain

$$\left\langle 0, \Delta E_w \left| \hat{\rho}_w \hat{H}_o \hat{\rho}_w \right| 0, \Delta E_w \right\rangle = \sum_{\substack{m\in band \\ all\, n}} e^2 \left| \left( \int \phi_n^\dagger \chi \phi_m d\vec{x} \right) \right|^2 (\lambda_n E_n - \lambda_m E_m) \tag{47}$$

where we have used

$$\sum_{\substack{m\in band \\ all\, n}} = \sum_{\substack{m\in band \\ n>0}} + \sum_{\substack{m\in band \\ n\in band}} + \sum_{\substack{m\in band \\ n<band}} \tag{48}$$

Next use (28) and the fact that $H_o$ is hermitian to obtain

$$(\lambda_n E_n - \lambda_m E_m)\left( \int \phi_n^\dagger \chi \phi_m d\vec{x} \right) = \int \phi_n^\dagger (H_o \chi - \chi H_o) \phi_m d\vec{x} = \int \phi_n^\dagger [H_o, \chi] \phi_m d\vec{x} \tag{49}$$

Use this and the relationship

$$[H_o, \chi] = -i\vec{\alpha} \cdot \vec{\nabla} \chi \tag{50}$$

in (47) to yield

$$\left\langle 0,\Delta E_w \left|\hat{\rho}_w \hat{H}_o \hat{\rho}_w\right| 0,\Delta E_w \right\rangle = -i \sum_{\substack{m\in \text{band}\\ \text{all } n}} e^2\left(\int \phi_m^\dagger(\vec{y})\chi(\vec{y})\phi_n(\vec{y})d\vec{y}\right)\left(\int \phi_n^\dagger(\vec{x})\left(\vec{\alpha}\cdot\vec{\nabla}\chi(\vec{x})\right)\phi_m(\vec{x})d\vec{x}\right) \tag{51}$$

Perform the summation over 'n' first and use (31) in the above to obtain

$$\left\langle 0,\Delta E_w \left|\hat{\rho}_w \hat{H}_o \hat{\rho}_w\right| 0,\Delta E_w \right\rangle = -i \sum_{m\in \text{band}} e^2\left(\int \phi_m^\dagger\left(\chi\vec{\alpha}\cdot\vec{\nabla}\chi\right)\phi_m d\vec{x}\right) = \frac{-i}{2}\sum_{m\in \text{band}} e^2\left(\int \phi_m^\dagger\left(\vec{\alpha}\cdot\vec{\nabla}\chi^2\right)\phi_m d\vec{x}\right) \tag{52}$$

Next, assume reasonable boundary conditions at $|\vec{x}| \to \infty$ and integrate by parts to yield

$$\left\langle 0,\Delta E_w \left|\hat{\rho}_w \hat{H}_o \hat{\rho}_w\right| 0,\Delta E_w \right\rangle = \frac{i}{2}\sum_{m\in \text{band}} e^2\left(\int \chi^2\vec{\nabla}\cdot\left(\phi_m^\dagger\vec{\alpha}\phi_m\right)d\vec{x}\right) \tag{53}$$

From the form of $\phi_m$ given in (30) we have that $\phi_m^\dagger\vec{\alpha}\phi_m = \text{constant}$ so that

$\vec{\nabla}\cdot\left(\phi_m^\dagger\vec{\alpha}\phi_m\right) = 0$, and therefore,

$$\left\langle 0,\Delta E_w \left|\hat{\rho}_w \hat{H}_o \hat{\rho}_w\right| 0,\Delta E_w \right\rangle = 0 \tag{54}$$

Therefore using the state vector $\left|0,\Delta E_w\right\rangle$ results in a mathematically consistent theory.

To determine $\left\langle 0\left|\hat{\rho}_w \hat{H}_o \hat{\rho}_w\right|0\right\rangle$ refer back to (42). In the first summation replace $m \in \text{band}$ with $m < 0$. Eliminate the second summation since no transitions corresponding to $n < \text{band}$ exist for the state vector $\left|0\right\rangle$. This results in

$$\left\langle 0\left|\hat{\rho}_w \hat{H}_o \hat{\rho}_w\right|0\right\rangle = \sum_{\substack{m<0\\ n>0}} e^2\left|\left(\int \phi_n^\dagger\chi\phi_m d\vec{x}\right)\right|^2\left(E_m + E_n\right) \tag{55}$$

This is, of course, greater than zero because each term in the sum is positive and not, in general, zero. What allows $\left\langle 0,\Delta E_w \left|\hat{\rho}_w \hat{H}_o \hat{\rho}_w\right| 0,\Delta E_w \right\rangle$ to equal zero is the second summation in (42), which is due to transitions to negative energy states underneath the band. This summation is negative because the term $(E_m - E_n)$ is negative. Therefore the second summation cancels out the first summation. These transitions to negative energy states underneath the band are necessary for a mathematically consistent theory.

In light of the above discussion the vacuum state will be redefined as follows. Let the vacuum be the state $|0,\Delta E_w\rangle$ in the limit that $\Delta E_w \to \infty$. This will be written as $|0,\Delta E_w \to \infty\rangle$. In calculations involving $|0,\Delta E_w \to \infty\rangle$, $\Delta E_w$ is assumed to be finite with the limit $\Delta E_w \to \infty$ taken at the end of the calculation. In the last section it was shown that $\langle 0,\Delta E_w|\hat{\rho}_w \hat{H}_o \hat{\rho}_w|0,\Delta E_w\rangle$ equals zero. This result does not change as $\Delta E_w \to \infty$. Therefore $\langle 0,\Delta E_w \to \infty|\hat{\rho}_w \hat{H}_o \hat{\rho}_w|0,\Delta E_w \to \infty\rangle$ equals zero which is the requirement for a mathematically consistent theory.

Now the vacuum state plays an important role in Dirac field theory and must satisfy certain requirements. First of all the vacuum must be a state where the charge and current are zero. In addition, it must be allow for the existence of antiparticles. The most important requirement of the vacuum is that it prevents the scattering of positive energy particles into negative energy states by perturbing electric fields. It is shown in Ref [9] that the states $|0\rangle$ and $|0,\Delta E_w \to \infty\rangle$ both meet these requirements.

The difference between $|0\rangle$ and $|0,\Delta E_w \to \infty\rangle$ may be understood by examining how the concept of infinity is handled in the expression "all negative energy states from $-m$ to $-\infty$ are occupied." For the state $|0,\Delta E_w \to \infty\rangle$ infinity is a limiting process in which the limit to infinity is taken at the end of any calculations. For the state $|0\rangle$ the limit to infinity is taken at the beginning of the calculation. Other than this $|0\rangle$ and $|0,\Delta E_W \to \infty\rangle$ are identical therefore we should expect that most tests of Dirac field theory such as scattering cross sections etc. would be identical in both cases.

## 4. Local charge conservation

As demonstrated in Appendix A the direct use the field operator anticommutator relationships (1) results in the Schwinger term $\left[\hat{\rho}(\vec{y}),\hat{\vec{J}}(\vec{x})\right]$ being equal to zero (see (6)). However when combined with the normal definition of the vacuum state $|0\rangle$ this results in the inconsistencies already discussed. To resolve these problems previous researchers have assumed that the "naive" use of (1) cannot be trusted and that the commutator $\left[\hat{\rho}(\vec{y}),\hat{\vec{J}}(\vec{x})\right]$ is not zero. In this and the next section we will show that the commutator equations (6) through (8) play an important role in local charge conservation and gauge invariance. It will be shown that the Schwinger term must be zero for local charge conservation and gauge invariance to hold.

In this section we will examine the requirements for local charge conservation. The time evolution of the state vector $|\Omega\rangle$ in the Schrodinger representation is

$$i\frac{\partial|\Omega\rangle}{\partial t} = \hat{H}|\Omega\rangle \; ; \qquad -i\frac{\partial\langle\Omega|}{\partial t} = \langle\Omega|\hat{H} \tag{56}$$

where the Hamiltonian operator is given by

$$\hat{H} = \hat{H}_0 + \int\left(-\hat{\vec{J}}(\vec{x})\cdot\vec{A}(\vec{x},t) + \hat{\rho}(\vec{x})A_0(\vec{x},t)\right)d\vec{x} \tag{57}$$

and where $(A_o, \vec{A})$ is the electric potential. In this discussion the electrical potential are taken to be unquantized, real valued, classical quantities. Note that natural units are used so that $\hbar = c = 1$. From (56) we have that

$$\frac{\partial\langle\Omega|\Omega\rangle}{\partial t} = 0 \tag{58}$$

The current and charge expectation values are defined by

$$\vec{J}_e = \frac{\langle\Omega|\hat{\vec{J}}|\Omega\rangle}{\langle\Omega|\Omega\rangle} \tag{59}$$

and

$$\rho_e = \frac{\langle\Omega|\hat{\rho}|\Omega\rangle}{\langle\Omega|\Omega\rangle} \tag{60}$$

In quantum mechanics the continuity equation can be written as

$$\frac{\partial\rho_e}{\partial t} = -\nabla\cdot\vec{J}_e \tag{61}$$

Use (56), (58), (59), and (60) in the above to obtain

$$i\langle\Omega|\left[\hat{H},\hat{\rho}\right]|\Omega\rangle = -\langle\Omega|\nabla\cdot\hat{\vec{J}}|\Omega\rangle \tag{62}$$

Next use (57) in the above to obtain

$$\langle\Omega|\left(\left[\hat{H}_o + \int\left(-\hat{\vec{J}}(\vec{y})\cdot\vec{A}(\vec{y},t) + \hat{\rho}(\vec{y})A_o(\vec{y},t)\right)d\vec{y}, \hat{\rho}(\vec{x})\right]\right)|\Omega\rangle = i\langle\Omega|\nabla\cdot\hat{\vec{J}}(\vec{x})|\Omega\rangle \tag{63}$$

Rearrange terms to obtain

$$\langle\Omega|\left(\left[\hat{H}_o,\hat{\rho}(\vec{x})\right] + \int\left(-\left[\hat{\vec{J}}(\vec{y}),\hat{\rho}(\vec{x})\right]\cdot\vec{A}(\vec{y},t) + \left[\hat{\rho}(\vec{y}),\hat{\rho}(\vec{x})\right]A_o(\vec{y},t)\right)d\vec{y}\right)|\Omega\rangle = i\langle\Omega|\nabla\cdot\hat{\vec{J}}(\vec{x})|\Omega\rangle \tag{64}$$

For this expression to be true for all possible combinations of $|\Omega\rangle$ and electric potential $\left(A_o, \vec{A}\right)$, the commutator relationships (6) through (8) must be true.

## 5. Gauge Invariance

In this section we shall examine the role that equations (6) through (8) play in the gauge invariance of Dirac field theory. The electromagnetic field is given in terms of the electric potential according to

$$\vec{E} = -\left(\frac{\partial \vec{A}}{\partial t} + \vec{\nabla} A_o\right); \quad \vec{B} = \vec{\nabla} \times \vec{A} \tag{65}$$

A change in the gauge is a change in the electric potential that produces no change in the electromagnetic field. Such a change is given by

$$\vec{A} \to \vec{A}' = \vec{A} - \vec{\nabla}\chi; \quad A_o \to A_o' = A_o + \frac{\partial \chi}{\partial t} \tag{66}$$

where $\chi(\vec{x}, t)$ is an arbitrary real valued function. Substitute the above expression into (57) to show that a gauge transformation transforms the Hamiltonian according to,

$$\hat{H}_g = \hat{H} + \int \left(\hat{\vec{J}} \cdot \vec{\nabla}\chi + \hat{\rho}\frac{\partial \chi}{\partial t}\right) d\vec{x} \tag{67}$$

If the original state vector $|\Omega\rangle$ satisfies (56) then the gauge transformed state vector $|\Omega_g\rangle$ satisfies

$$i\frac{\partial |\Omega_g\rangle}{\partial t} = \hat{H}_g |\Omega_g\rangle = \left(\hat{H} + \int \left(\hat{\vec{J}} \cdot \vec{\nabla}\chi + \hat{\rho}\frac{\partial \chi}{\partial t}\right) d\vec{x}\right)|\Omega_g\rangle \tag{68}$$

It will be shown that $|\Omega\rangle$ and the gauge transformed state $|\Omega_g\rangle$ are related by

$$\left|\Omega_g\right\rangle = e^{-i\hat{\rho}_w(t)}\left|\Omega\right\rangle \tag{69}$$

where

$$\hat{\rho}_W(t) = \int \hat{\rho}(\vec{x})\chi(\vec{x},t)d\vec{x} \tag{70}$$

(Note that this expression is the same as (9) except that $\chi$ is time dependent). To prove this assume that $\left|\Omega\right\rangle$ satisfies (56) and substitute $\left|\Omega_g\right\rangle = e^{-i\hat{\rho}_w(t)}\left|\Omega\right\rangle$ into (68) to obtain

$$\frac{\partial\hat{\rho}_W}{\partial t}e^{-i\hat{\rho}_w}\left|\Omega\right\rangle + ie^{-i\hat{\rho}_w}\frac{\partial\left|\Omega\right\rangle}{\partial t} = \left[\hat{H} + \int\left(\hat{\vec{J}}\cdot\vec{\nabla}\chi + \hat{\rho}\frac{\partial\chi}{\partial t}\right)d\vec{x}\right]e^{-i\hat{\rho}_w}\left|\Omega\right\rangle \tag{71}$$

From ( 70) we have that

$$\frac{\partial\rho_W(t)}{\partial t} = \int\hat{\rho}(\vec{x})\frac{\partial\chi(\vec{x},t)}{\partial t}d\vec{x} \tag{72}$$

Use this in (71) to yield

$$ie^{-i\hat{\rho}_w}\frac{\partial\left|\Omega\right\rangle}{\partial t} = \left[\hat{H} + \int\hat{\vec{J}}\cdot\vec{\nabla}\chi d\vec{x}\right]e^{-i\hat{\rho}_w}\left|\Omega\right\rangle \tag{73}$$

From (6) and (7) we have that,

$$\left[e^{-i\hat{\rho}_w},\hat{\vec{J}}(\vec{x})\right] = 0 \tag{74}$$

and

$$\left[e^{-i\hat{\rho}_w},\hat{\rho}(\vec{x})\right] = 0 \tag{75}$$

Using (10) and (12) it is shown in Appendix B that

$$\left[\hat{H}_o, e^{-i\hat{\rho}_w}\right] = -e^{-i\hat{\rho}_w}\int\hat{\vec{J}}\cdot\vec{\nabla}\chi d\vec{x} \tag{76}$$

Next use the definition of $\hat{H}$, (57), and the above commutator relations to obtain

$$\hat{H}e^{-i\hat{\rho}_w}|\Omega\rangle = \left[\hat{H}_o + \int\left(-\hat{\vec{J}}\cdot\vec{A} + \hat{\rho}A_o\right)d\vec{x}\right]e^{-i\hat{\rho}_w}|\Omega\rangle$$
$$= e^{-i\hat{\rho}_w}\left[\hat{H}_o - \int\hat{\vec{J}}\cdot\vec{\nabla}\chi\right]|\Omega\rangle + e^{-i\hat{\rho}_w}\int\left(-\hat{\vec{J}}\cdot\vec{A} + \hat{\rho}A_o\right)d\vec{x}|\Omega\rangle \qquad (77)$$
$$= \left[e^{-i\hat{\rho}_w}\hat{H} - \left(\int\hat{\vec{J}}\cdot\vec{\nabla}\chi d\vec{x}\right)e^{-i\hat{\rho}_w}\right]|\Omega\rangle$$

Use this in (73) to obtain

$$ie^{-i\hat{\rho}_w}\frac{\partial|\Omega\rangle}{\partial t} = e^{-i\hat{\rho}_w}\hat{H}|\Omega\rangle \qquad (78)$$

Now operate on both sides with $e^{i\hat{\rho}_w}$ to recover (56).

If quantum field theory is gauge invariant then a gauge transformation cannot produce a change in any physical observable. This includes the current and charge expectation value. For the current expectation value the following equation must be true

$$\frac{\langle\Omega_g|\vec{J}(\vec{x})|\Omega_g\rangle}{\langle\Omega_g|\Omega_g\rangle} = \frac{\langle\Omega|\vec{J}(\vec{x})|\Omega\rangle}{\langle\Omega|\Omega\rangle} \qquad (79)$$

From (69) we have that

$$\langle\Omega_g|\Omega_g\rangle = \langle\Omega|e^{+i\hat{\rho}_w(t)}e^{-i\hat{\rho}_w(t)}|\Omega\rangle = \langle\Omega|\Omega\rangle \qquad (80)$$

Use this in (79) to obtain

$$\langle\Omega_g|\vec{J}(\vec{x})|\Omega_g\rangle = \langle\Omega|e^{+i\hat{\rho}_w(t)}\vec{J}(\vec{x})e^{-i\hat{\rho}_w(t)}|\Omega\rangle = \langle\Omega|\vec{J}(\vec{x})|\Omega\rangle \qquad (81)$$

where (74) has been used. A similar relationship can be shown to hold for the charge expectation value. Therefore we have shown that Dirac field theory is gauge invariant. The proof utilized (6) through (8) and (56).

## 6. Perturbation theory and the Schwinger term

The purpose of the last two sections was to highlight the importance of the commutators in equations (6) through (8) in Dirac field theory. It was shown that the physical laws of local charge conservation and gauge invariance follow from the field operator anticommutator relationship (1) and the evolution equation for the state vector (56). In particular we are interested in the commutator $\left[\hat{\rho}(\vec{x}),\hat{\vec{J}}(\vec{y})\right]$. The requirements of local charge conservation and gauge invariance require that this quantity be equal to zero.

As discussed in the introduction the problems associated with gauge invariance occur when the vacuum current is calculated using perturbation theory. In this section we will examine the important role that the Schwinger term plays in this problem. It will be shown that if the Schwinger term is not zero then the change in the vacuum current due to an applied external potential cannot be gauge invariant.

From equation 8.3 of Pauli [7] the first order change in the vacuum current due to an external perturbing electric potential is given by

$$\vec{J}^{(1)}_{vac}(\vec{x},t) = i\langle 0\left|\left[\hat{\vec{J}}(\vec{x},t),\int d\vec{y}\int_{-\infty}^{t} dt'\left(-\hat{\vec{J}}(\vec{y},t')\cdot\vec{A}(\vec{y},t')+\hat{\rho}(\vec{y},t')A_0(\vec{y},t')\right)\right]\right|0\rangle \quad (82)$$

In the above expression the operators $\hat{\vec{J}}(\vec{x},t)$ and $\hat{\rho}(\vec{x},t)$ are the current and charge operators, respectively, in the interaction representation. They are related to the Schrodinger operators by

$$\hat{\vec{J}}(\vec{x},t) = e^{i\hat{H}_0 t}\hat{\vec{J}}(\vec{x})e^{-i\hat{H}_0 t} \text{ and } \hat{\rho}(\vec{x},t) = e^{i\hat{H}_0 t}\hat{\rho}(\vec{x})e^{-i\hat{H}_0 t} \quad (83)$$

According to equation 3.11 of Pauli [7] the above interaction operators satisfy

$$\frac{\partial \hat{\rho}(\vec{x},t)}{\partial t} = -\vec{\nabla}\cdot\hat{\vec{J}}(\vec{x},t) \tag{84}$$

The change in the vacuum current $\delta\vec{J}^{(1)}_{vac}(\vec{x},t)$ due to a gauge transformation is obtained by using (66) in (82) and separating out the gauge dependent part to yield

$$\delta\vec{J}^{(1)}_{vac}(\vec{x},t) = i\langle 0 \| \left[ \hat{\vec{J}}(\vec{x},t), \int d\vec{y} \int_{-\infty}^{t} dt' \left( \hat{\vec{J}}(\vec{y},t')\cdot\vec{\nabla}\chi(\vec{y},t') + \hat{\rho}(\vec{y},t')\frac{\partial\chi(\vec{y},t')}{\partial t'} \right) \right] \| 0 \rangle \tag{85}$$

If quantum field theory is gauge invariant then a gauge transformation of the electric potential should produce no change in any observable quantity. Therefore $\delta\vec{J}^{(1)}_{vac}(\vec{x},t)$ should be zero. To determine if this is the case we will solve the above equation as follows.

$$\int_{-\infty}^{t} dt'\hat{\rho}(\vec{y},t')\frac{\partial\chi(\vec{y},t')}{\partial t'} = \Big|_{-\infty}^{t} \hat{\rho}(\vec{y},t')\chi(\vec{y},t') - \int_{-\infty}^{t} dt'\chi(\vec{y},t')\frac{\partial\hat{\rho}(\vec{y},t')}{\partial t'} \tag{86}$$

Assume that $\chi(\vec{y},t) = 0$ at $t \to -\infty$. Use this and (84) in the above expression to obtain

$$\int_{-\infty}^{t} dt'\hat{\rho}(\vec{y},t')\frac{\partial\chi(\vec{y},t')}{\partial t'} = \hat{\rho}(\vec{y},t)\chi(\vec{y},t) + \int_{-\infty}^{t} dt'\chi(\vec{y},t')\vec{\nabla}\cdot\hat{\vec{J}}(\vec{y},t') \tag{87}$$

Substitute this into (85) to obtain

$$\begin{aligned}\delta\vec{J}^{(1)}_{vac}(\vec{x},t) = {} & i\langle 0 \| \left[ \hat{\vec{J}}(\vec{x},t), \int d\vec{y} \int_{-\infty}^{t} dt' \left( \hat{\vec{J}}(\vec{y},t')\cdot\vec{\nabla}\chi(\vec{y},t') + \chi(\vec{y},t')\vec{\nabla}\cdot\hat{\vec{J}}(\vec{y},t') \right) \right] \| 0 \rangle \\ & + i\langle 0 \| \left[ \hat{\vec{J}}(\vec{x},t), \int \hat{\rho}(\vec{y},t)\chi(\vec{y},t)d\vec{y} \right] \| 0 \rangle \end{aligned} \tag{88}$$

Rearrange terms to obtain

$$\delta\vec{J}^{(1)}_{vac}(\vec{x},t) = i\left\langle 0\left|\left[\hat{\vec{J}}(\vec{x},t), \int_{-\infty}^{t} dt'\int d\vec{y}\vec{\nabla}\cdot\left(\hat{\vec{J}}(\vec{y},t')\chi(\vec{y},t')\right)\right]\right|0\right\rangle + i\left\langle 0\left|\left[\hat{\vec{J}}(\vec{x},t), \int\hat{\rho}(\vec{y},t)\chi(\vec{y},t)d\vec{y}\right]\right|0\right\rangle \quad (89)$$

Assume reasonable boundary conditions at $|\vec{y}| \to \infty$ so that

$$\int d\vec{y}\vec{\nabla}\cdot\left(\hat{\vec{J}}(\vec{y},t')\chi(\vec{y},t')\right) = 0 \quad (90)$$

Use this to obtain

$$\delta\vec{J}^{(1)}_{vac}(\vec{x},t) = i\left\langle 0\left|\left[\hat{\vec{J}}(\vec{x},t), \int\hat{\rho}(\vec{y},t)\chi(\vec{y},t)d\vec{y}\right]\right|0\right\rangle = i\int\left\langle 0\left|\left[\hat{\vec{J}}(\vec{x},t),\hat{\rho}(\vec{y},t)\right]\right|0\right\rangle\chi(\vec{y},t)d\vec{y} \quad (91)$$

Use (83) and the fact that $\hat{H}_o|0\rangle = 0$ in the above to obtain

$$\delta\vec{J}^{(1)}_{vac}(\vec{x},t) = i\int\left\langle 0\left|\left[\hat{\vec{J}}(\vec{x}),\hat{\rho}(\vec{y})\right]\right|0\right\rangle\chi(\vec{y},t)d\vec{y} = i\left\langle 0\left|\left[\hat{\vec{J}}(\vec{x}),\hat{\rho}_w\right]\right|0\right\rangle \quad (92)$$

where $\hat{\rho}_w$ is given by ( 70). Therefore for $\delta\vec{J}^{(1)}_{vac}(\vec{x},t)$ to be zero, for arbitrary $\chi(\vec{y},t)$, the Schwinger term $\left\langle 0\left|\left[\hat{\vec{J}}(\vec{x}),\hat{\rho}(\vec{y})\right]\right|0\right\rangle$ (or $\left\langle 0\left|\left[\hat{\vec{J}}(\vec{x}),\hat{\rho}_w\right]\right|0\right\rangle$) must be zero. The reason why calculations of the vacuum current always yield non-gauge invariant terms is because when $|0\rangle$ is used as the vacuum state the Schwinger term $\left\langle 0\left|\left[\hat{\vec{J}}(\vec{x}),\hat{\rho}(\vec{y})\right]\right|0\right\rangle$ is not zero. This will be demonstrated in the next section.

## 7. Calculation of the Gauge Current I

It was mentioned in the Introduction that non-gauge invariant terms appear in a calculation of the vacuum current (see [6], [7], [1], and [8]). One problem with trying to understand why these terms appear in a theory that is suppose to be gauge invariant, is that these terms are divergent. These leads to the speculation that these terms are somehow an artifact of the mathematics.

Now why do divergent terms arise in the calculation of the vacuum current? The reason is that in performing this calculation integrals of the form $\int f(\vec{p})d\vec{p}$ arise. In 3-1D space-time $d\vec{p} \rightarrow p^2 dp$ where $p = |\vec{p}|$. The $p^2$ term makes the integral divergent. Therefore it is instructive to work this problem in 1-1D space-time. In this case $d\vec{p} \rightarrow dp$ and the resulting integrals will be finite.

We shall, then, calculate $\delta\vec{J}_{vac}^{(1)}(\vec{x},t)$ in 1-1D space for a specific gauge transformation. The result will be finite and well defined. No ultra-violet cutoff or other mathematical devices are needed to obtain a reasonable result. It will be shown that when $|0\rangle$ is used as the vacuum state then $\delta\vec{J}_{vac}^{(1)}(\vec{x},t)$ is not zero so that the theory is not gauge invariant. In the next section the calculation shall be redone using $|0, \Delta E_w \rightarrow \infty\rangle$ and it will be shown that in this case $\delta\vec{J}_{vac}^{(1)}(\vec{x},t)$ is zero and the theory is gauge invariant.

In 1-1D space, where z is chosen as the space dimension, the eigenfunction solutions to (28) are

$$\phi_n(z) = u_n e^{ip_n z} \tag{93}$$

$$u_n = \sqrt{\frac{\lambda_n E_n + m}{2\lambda_n E_n L}} \begin{pmatrix} 1 \\ 0 \\ p_n / (\lambda_n E_n + m) \\ 0 \end{pmatrix} \quad \text{for spin} = 1 \tag{94}$$

$$u_n = \sqrt{\frac{\lambda_n E_n + m}{2\lambda_n E_n L}} \begin{pmatrix} 0 \\ 1 \\ 0 \\ -p_n / (\lambda_n E_n + m) \end{pmatrix} \quad \text{for spin} = 2 \tag{95}$$

where L is the normalization length, i.e., $\phi_n(z)$ is normalized to a probability of 1 in a region of length L where $L\rightarrow\infty$. Also periodic boundary conditions are assumed so that $\phi_n(z) = \phi_n(z+L)$. Therefore the momentum takes on the discrete values $p_n=2\pi n/L$ where n is an integer. The density of states in the momentum range $\Delta p$ is $\Delta n=(L/2\pi)\Delta p$. Use (26) in (2) to obtain for the current operator

$$\hat{J}(z) = \frac{e}{2}\left\{\sum_{m,n} \phi_n^\dagger \alpha_z \phi_m \left(\hat{a}_n^\dagger \hat{a}_m - \hat{a}_m \hat{a}_n^\dagger\right)\right\} = e\left\{\sum_{m,n}\left(\phi_n^\dagger \alpha_z \phi_m\right)\left(\hat{a}_n^\dagger \hat{a}_m - \frac{\delta_{mn}}{2}\right)\right\} \tag{96}$$

where

$$\alpha_z = \begin{pmatrix} 0 & 0 & 1 & 0 \\ 0 & 0 & 0 & -1 \\ 1 & 0 & 0 & 0 \\ 0 & -1 & 0 & 0 \end{pmatrix} \tag{97}$$

Operate on the vacuum state $|0\rangle$ to obtain

$$\hat{J}(z)|0\rangle = e\sum_{\substack{m<0 \\ n>0}}\left(\phi_n^\dagger \alpha_z \phi_m\right)\hat{a}_n^\dagger \hat{a}_m|0\rangle + J_{vac}|0\rangle \tag{98}$$

where

$$J_{vac} = \langle 0|\hat{J}(z)|0\rangle \tag{99}$$

$J_{vac}$ is a constant that we can normally set equal to zero. Similarly refer to (38) to obtain

$$\hat{\rho}_w|0\rangle = e\sum_{\substack{m<0\\n>0}}\left(\int \phi_n^\dagger \chi \phi_m dz\right)\hat{a}_n^\dagger \hat{a}_m|0\rangle + \rho_{w,vac}|0\rangle \tag{100}$$

where $\rho_{w,vac}$ is a constant given by

$$\rho_{w,vac} = \langle 0|\hat{\rho}_w|0\rangle \tag{101}$$

Use the above equations to obtain in 1-1D space

$$\langle 0|\left[\hat{J}(z),\hat{\rho}_w\right]|0\rangle = e^2\sum_{\substack{m<0\\n>0}}\left\{\left(\phi_m^\dagger(z)\alpha_z\phi_n(z)\right)\left(\int_{-L/2}^{+L/2}\phi_n^\dagger\chi\phi_m dz\right)\right\} - \text{h.c.} \tag{102}$$

where h.c. means to take the hermitian conjugate of the proceeding term. Define

$$I(|0\rangle) \equiv \sum_{\substack{m<0\\n>0}}\left\{\left(\phi_m^\dagger(z)\alpha_z\phi_n(z)\right)\left(\int_{-L/2}^{+L/2}\phi_n^\dagger\chi\phi_m dz\right)\right\} \tag{103}$$

Therefore

$$\langle 0|\left[\hat{J}(z),\hat{\rho}_w\right]|0\rangle = e^2\left(I(|0\rangle) - \text{h.c.}\right) \tag{104}$$

Use (93) in the above to obtain

$$I(|0\rangle) = \sum_{\substack{m<0\\n>0}}\left\{\left(\left(u_m^\dagger\alpha_z u_n\right)\left(u_n^\dagger u_m\right)e^{i(p_n-p_m)z}\right)\left(\int_{-L/2}^{+L/2}\chi e^{-i(p_n-p_m)z}\right)\right\} \tag{105}$$

From (94) and (95) after some algebra it can be shown that

$$\left(u_m^\dagger \alpha_z u_n\right)\left(u_n^\dagger u_m\right) = \left\langle \begin{array}{l} \frac{1}{2L^2}\left(\frac{p_n}{\lambda_n E_n} + \frac{p_m}{\lambda_m E_m}\right) \text{ if n and m have the same spin.} \\ \quad 0 \text{ if n and m have different spin.} \end{array} \right. \tag{106}$$

Substitute this into (105) to obtain

$$I(|0\rangle) = \frac{1}{2L^2} \sum_{\substack{m<0 \\ n>0}} \left\{ \left( s_{nm} \left( \frac{p_n}{E_n} - \frac{p_m}{E_m} \right) e^{i(p_n - p_m)z} \right) \left( \int_{-L/2}^{+L/2} \chi e^{-i(p_n - p_m)z} dz \right) \right\} \tag{107}$$

where

$$s_{nm} = \left\langle \begin{array}{l} 1 \text{ if spin of n = spin of m} \\ 0 \text{ if the spins are not equal} \end{array} \right. \tag{108}$$

Now replace the summation sign with an integral according to

$$\sum_{\substack{m<0 \\ n>0}} \to 2 \int_{-\infty}^{+\infty} \frac{L}{2\pi} dp_m \int_{-\infty}^{+\infty} \frac{L}{2\pi} dp_n \tag{109}$$

where the leading factor of 2 is due to summation over the two spin states. This yields

$$I(|0\rangle) = \frac{1}{4\pi^2} \int_{-\infty}^{+\infty} dp_m \int_{-\infty}^{+\infty} dp_n \left\{ \left( \left( \frac{p_n}{E_n} - \frac{p_m}{E_m} \right) e^{i(p_n - p_m)z} \right) \left( \int_{-L/2}^{+L/2} \chi e^{-i(p_n - p_m)z} dz \right) \right\}$$

(110)

We shall solve the above for a paticular $\chi(z,t)$. Let $\chi(z,t)$ be given by

$$\chi(z,t) = V_o(t)\cos(kz) = \frac{V_o(t)}{2}\left(e^{ikz} + e^{-ikz}\right) \tag{111}$$

where $V_o(t) = 0$ at $t \to -\infty$.

Take $L \to \infty$ in the above integral to obtain

$$I(|0\rangle)=\frac{V_o}{8\pi^2}\int_{-\infty}^{+\infty}dp_m\int_{-\infty}^{+\infty}dp_n\left\{\left(\left(\frac{p_n}{E_n}-\frac{p_m}{E_m}\right)e^{i(p_n-p_m)z}\right)\begin{pmatrix}\delta(p_n-p_m-k)\\+\delta(p_n-p_m+k)\end{pmatrix}\right\} \quad (112)$$

This yields

$$I(|0\rangle)=\frac{V_o}{8\pi^2}\int_{-\infty}^{+\infty}dp_m\left(\left(\frac{p_m+k}{\sqrt{(p_m+k)^2+m^2}}-\frac{p_m}{\sqrt{p_m^2+m^2}}\right)e^{ikz}\right)+(k\to -k) \quad (113)$$

where (k→–k) means to repeat the term to the left but substitute –k for k.

Now evaluate the integral as follows.

$$\int_{-\infty}^{+\infty}dp\left(\frac{p+k}{\sqrt{(p+k)^2+m^2}}-\frac{p}{\sqrt{p^2+m^2}}\right)\underset{r\to\infty}{=}\frac{1}{2}\int_{-r}^{+r}d\left(\sqrt{(p+k)^2+m^2}-\sqrt{p^2+m^2}\right)$$

$$\underset{r\to\infty}{=}\frac{1}{2}\Big|_{-r}^{+r}\left(\sqrt{(p+k)^2+m^2}-\sqrt{p^2+m^2}\right) \quad (114)$$

$$\underset{r\to\infty}{=}\frac{1}{2}\left(\sqrt{(r+k)^2+m^2}-\sqrt{(r-k)^2+m^2}\right)$$

Now

$$\sqrt{(r\pm k)^2+m^2}=\left(\sqrt{r^2\pm 2rk+k^2+m^2}\right)\underset{r\to\infty}{=}r\left(1\pm\frac{k}{r}+O\left(\frac{1}{r^2}\right)\right) \quad (115)$$

Use this in (114) to obtain

$$\int_{-\infty}^{+\infty}dp\left(\frac{p+k}{\sqrt{(p+k)^2+m^2}}-\frac{p}{\sqrt{p^2+m^2}}\right)=k \quad (116)$$

Use this result in (113) to obtain

$$I(|0\rangle)=\frac{iV_ok}{4\pi^2}\sin(kz) \quad (117)$$

Use this in (104) to obtain

$$\langle 0|\left[\hat{J}(z),\hat{\rho}_w\right]|0\rangle = i\frac{e^2 V_o k}{2\pi^2}\sin(kz) \tag{118}$$

Therefore the Schwinger term is not zero. This leads to a non-gauge invariant result for the vacuum current which can be shown by substituting the above into (92) to obtain

$$\delta\vec{J}^{(1)}_{vac}(z,t) = -\frac{e^2 V_o k}{2\pi^2}\sin(kz) \tag{119}$$

The fact that this result is not zero shows that the first order change in the current for $|0\rangle$ is not gauge invariant. Note that this result is perfectly consistent with other discussions of the vacuum current (See Refs. [6], [7], [1], and [8]). In all these cases the calculation of the vacuum current contains non-gauge invariant terms which need to be dropped to produce a gauge invariant result. The difference here is that the calculation is done in 1-1D space-time so that the result is finite. The reason, then, for this non-gauge invariant result is not due to some unknown mathematical artifact but due to the fact that theory is not gauge invariant from the start when $|0\rangle$ is used as the vacuum state.

## 8. Calculation of Gauge Current II

In this section we shall calculate $\delta\vec{J}^{(1)}_{vac}(z,t)$ for the case where the state vector $|0,\Delta E_w \to \infty\rangle$ is used as the vacuum state. In this case (92) becomes

$$\delta\vec{J}^{(1)}_{vac}(z,t) = i\langle 0,\Delta E_w \to \infty|\left[\hat{\vec{J}}(z),\hat{\rho}_w\right]|0,\Delta E_w \to \infty\rangle \tag{120}$$

Use (96) and the definition of $|0,\Delta E_w\rangle$ to obtain

$$\hat{J}(z)\left|0,\Delta E_w\right\rangle = e\sum_{\substack{m\in band\\ n>0}}\left(\phi_n^{\dagger}\alpha_z\phi_m\right)\hat{a}_n^{\dagger}\hat{a}_m\left|0,\Delta E_w\right\rangle + e\sum_{\substack{m\in band\\ n<band}}\left(\phi_n^{\dagger}\alpha_z\phi_m\right)\hat{a}_n^{\dagger}\hat{a}_m\left|0,\Delta E_w\right\rangle + J'_{vac}\left|0,\Delta E_w\right\rangle \quad (121)$$

where $J'_{vac}$ is a constant given by

$$J'_{vac} = \left\langle 0,\Delta E_w\right|\hat{J}(z)\left|0,\Delta E_w\right\rangle \quad (122)$$

From Eq. (39) we obtain

$$\hat{\rho}_w\left|0,\Delta E_w\right\rangle = e\sum_{\substack{m\in band\\ n>0}}\left(\int\phi_n^{\dagger}\chi\phi_m dz\right)\hat{a}_n^{\dagger}\hat{a}_m\left|0,\Delta E_w\right\rangle + e\sum_{\substack{m\in band\\ n<band}}\left(\int\phi_n^{\dagger}\chi\phi_m dz\right)\hat{a}_n^{\dagger}\hat{a}_m\left|0,\Delta E_w\right\rangle + \rho'_{w,vac}\left|0,\Delta E_w\right\rangle \quad (123)$$

From the above equations it can be shown that

$$\left\langle 0,\Delta E_w\right|\left[\hat{J}(z),\hat{\rho}_w\right]\left|0,\Delta E_w\right\rangle = e^2\left(I_{(+)}\left(\left|0,\Delta E_w\right\rangle\right) + I_{(-)}\left(\left|0,\Delta E_w\right\rangle\right)\right) - h.c. \quad (124)$$

where

$$I_{(+)}\left(\left|0,\Delta E_w\right\rangle\right) = \frac{1}{2L^2}\sum_{\substack{m\in band\\ n>0}}\left\{\left(s_{nm}\left(\frac{p_n}{E_n}-\frac{p_m}{E_m}\right)e^{i(p_n-p_m)z}\right)\left(\int_{-L/2}^{+L/2}\chi e^{-i(p_n-p_m)z}dz\right)\right\} \quad (125)$$

and

$$I_{(-)}\left(\left|0,\Delta E_w\right\rangle\right) = \frac{1}{2L^2}\sum_{\substack{m\in band\\ n<band}}\left\{\left(s_{nm}\left(-\frac{p_n}{E_n}-\frac{p_m}{E_m}\right)e^{i(p_n-p_m)z}\right)\left(\int_{-L/2}^{+L/2}\chi e^{-i(p_n-p_m)z}dz\right)\right\} \quad (126)$$

For the quantum states $m \in$ band the energy is between $-m$ and $-(m+\Delta E_w)$. Let the momentum $p_m$ range from $-r$ to $+r$ where $\sqrt{r^2 + m^2} = m + \Delta E_w$. Therefore

$$\sum_{\substack{m \in \text{band} \\ n>0}} \to 2\int_{-r}^{+r} \frac{L}{2\pi} dp_m \int_{-\infty}^{+\infty} \frac{L}{2\pi} dp_n \tag{127}$$

Use this and (111) for $\chi$ and let $L \to \infty$ to obtain

$$I_{(+)}\left(\left|0, \Delta E_w\right\rangle\right) = \frac{V_o}{8\pi^2} \int_{-r}^{+r} dp_m \int_{-\infty}^{+\infty} dp_n \left\{\left(\left(\frac{p_n}{E_n} - \frac{p_m}{E_m}\right) e^{i(p_n - p_m)z}\right)\begin{pmatrix} \delta(p_n - p_m - k) \\ + \delta(p_n - p_m + k) \end{pmatrix}\right\} \tag{128}$$

In the limit $\Delta E_w \to \infty$ (so that $r \to \infty$) we obtain (see (112))

$$I_{(+)}\left(\left|0, \Delta E_w \to \infty\right\rangle\right) = I\left(\left|0\right\rangle\right) \tag{129}$$

Next evaluate $I_{(-)}\left(\left|0, \Delta E_w\right\rangle\right)$. For n<band the energy of the state n is less than $-(m + \Delta E_w)$ and the momentum $p_n$ is within the range $+r$ to $+\infty$ or $-r$ to $-\infty$.

Therefore

$$\sum_{\substack{m \in \text{band} \\ n < \text{band}}} \to 2\int_{-r}^{+r} \frac{L}{2\pi} dp_m \left[\left(\int_{+r}^{+\infty} \frac{L}{2\pi} dp_n\right) + \left(\int_{-\infty}^{-r} \frac{L}{2\pi} dp_n\right)\right] \tag{130}$$

Use this and (111) for $\chi$ in (126) and let $L \to \infty$ to obtain

$$I_{(-)}\left(\left|0, \Delta E_w\right\rangle\right) = \frac{V_o}{8\pi^2} \int_{-r}^{+r} dp_m \left[\left(\int_{+r}^{+\infty} dp_n\right) + \left(\int_{-\infty}^{-r} dp_n\right)\right] g(p_n, p_m) \tag{131}$$

where

$$g(p_n,p_m) \equiv f(p_n,p_m)\begin{pmatrix}\delta(p_n - p_m - k)\\ +\delta(p_n - p_m + k)\end{pmatrix} \tag{132}$$

and

$$f(p_n,p_m) \equiv \left(-\frac{p_n}{E_n} - \frac{p_m}{E_m}\right)e^{i(p_n - p_m)z} \tag{133}$$

Now assume that k>0 and r>k (eventually we will have $r \to \infty$ ) so that

$$\int_{-r}^{+r} dp_m \int_{+r}^{+\infty} dp_n f(p_n,p_m)\delta(p_n - p_m + k) = 0 \tag{134}$$

$$\int_{-r}^{+r} dp_m \int_{+r}^{+\infty} dp_n f(p_n,p_m)\delta(p_n - p_m - k) = \int_{r-k}^{r} f(p_m + k, p_m)dp_m \tag{135}$$

$$\int_{-r}^{+r} dp_m \int_{-\infty}^{-r} dp_n f(p_n,p_m)\delta(p_n - p_m - k) = 0 \tag{136}$$

$$\int_{-r}^{+r} dp_m \int_{-\infty}^{-r} dp_n f(p_n,p_m)\delta(p_n - p_m + k) = \int_{-r}^{-r+k} f(p_m - k, p_m)dp_m \tag{137}$$

Therefore

$$I_{(-)}\left(\left|0,\Delta E_w\right\rangle\right) = \left(-\frac{V_o}{8\pi^2}\right)\int_{-r}^{-r+k} dp\left(\frac{p-k}{\sqrt{(p-k)^2 + m^2}} + \frac{p}{\sqrt{p^2 + m^2}}\right)e^{-ikz}$$
$$+ \left(-\frac{V_o}{8\pi^2}\right)\int_{r-k}^{r} dp\left(\frac{p+k}{\sqrt{(p+k)^2 + m^2}} + \frac{p}{\sqrt{p^2 + m^2}}\right)e^{ikz}$$

(138)

$$I_{(-)}\left(\left|0,\Delta E_w\right\rangle\right)=\left(-\frac{V_o}{8\pi^2}\right)\frac{1}{2}\Bigg|_{-r}^{-r+k}\left(\sqrt{(p-k)^2+m^2}+\sqrt{p^2+m^2}\right)e^{-ikz}$$
$$+\left(-\frac{V_o}{8\pi^2}\right)\frac{1}{2}\Bigg|_{r-k}^{r}\left(\sqrt{(p+k)^2+m^2}+\sqrt{p^2+m^2}\right)e^{ikz}$$

(139)

$$I_{(-)}\left(\left|0,\Delta E_w\right\rangle\right)=\left(-\frac{V_o}{16\pi^2}\right)\left(\sqrt{(r-k)^2+m^2}-\sqrt{(r+k)^2+m^2}\right)e^{-ikz}$$
$$+\left(-\frac{V_o}{16\pi^2}\right)\left(\sqrt{(r+k)^2+m^2}-\sqrt{(r-k)^2+m^2}\right)e^{ikz}$$

(140)

Take the limit $\Delta E_w \to \infty$ (and $r \to \infty$) and use (115) in the above and refer to (117) to obtain

$$I_{(-)}\left(\left|0,\Delta E_w \to \infty\right\rangle\right)=-I\left(\left|0\right\rangle\right) \tag{141}$$

Use this and (129) in (124) to show that

$$\left\langle 0,\Delta E_w \to 0\right|\left[\hat{\vec{J}}(z),\hat{\rho}_w\right]\left|0,\Delta E_w \to 0\right\rangle = 0 \tag{142}$$

So that the Schwinger term is zero. Use this in (120) to obtain

$$\delta\vec{J}^{(1)}_{vac}(z)=0 \tag{143}$$

Therefore the first order change in the current due to a gauge transformation is zero for the state $\left|0,\Delta E_w \to 0\right\rangle$, so that if $\left|0,\Delta E_w \to 0\right\rangle$ is used as the vacuum state then Dirac field theory will be gauge invariant.

Now what is the difference between $\left|0\right\rangle$ and $\left|0,\Delta E_w \to 0\right\rangle$ that accounts for these results? In calculating $\left\langle 0\right|\left[\hat{J}(z),\hat{\rho}_w\right]\left|0\right\rangle$ in Section 7 we have to determine the quantity $I(|0\rangle)$. $I(|0\rangle)$ may be thought of as being due to transitions from negative energy states to positive energy states. In calculating $\langle 0, \Delta E_w \to \infty | [\hat{\vec{J}}(z), \hat{\rho}_w] | 0, \Delta E_w \to \infty \rangle$ in Section 8 we have to determine two quantities $I_{(+)}(|0, \Delta E_w\rangle)$ and $I_{(-)}(|0, \Delta E_w\rangle)$. $I_{(+)}(|0, \Delta E_w\rangle)$ may be thought of as being due to transitions from the negative energy band to positive energy states. $I_{(+)}(|0, \Delta E_w\rangle)$, then, corresbonds to $I(|0\rangle)$ and in the limit that $\Delta E_w \to \infty$ these two quantities will be equal. $I_{(-)}(|0, \Delta E_w\rangle)$, on the other hand is a new term. It corresponds to transitions from the negative energy band to negative energy states underneath the negative energy band. In the limit that $\Delta E_w \to \infty$ this term does not go to zero. It cancels out $I_{(+)}(|0, \Delta E_w\rangle)$ so that $\langle 0, \Delta E_w \to \infty | [\hat{\vec{J}}(z), \hat{\rho}_w] | 0, \Delta E_w \to \infty \rangle = 0$ and the theory is gauge invariant.

The above discussion should provide some insight into why the direct calculation of the vacuum current by other researchers does not give a gauge invariant result. The reason for this is that when the mathematical calculations in these works are examined it is apparent that only transitions from negative energy states to positive energy states are included. This is because the vacuum state is $|0\rangle$ and these are the only transitions allowed. As we have just seen this leads to a non-gauge invariant result. To obtain a gauge invariant result the vacuum state must be $|0, \Delta E_w \to \infty\rangle$ which allows transitions from the negative energy band to states underneath the negative energy band in the limit that $\Delta E_w \to \infty$.

## 9. Discussion and Summary

In this article we have examined the role of the Schwinger term in Dirac field theory. A calculation of $\left[\hat{\rho}(\vec{x}),\hat{\vec{J}}(\vec{y})\right]$ using the anticommutator relations (1) shows that this quantity is zero. In addition a formal analysis shows that the commutator $\left[\hat{\rho}(\vec{x}),\hat{\vec{J}}(\vec{y})\right]$ must be zero for Dirac theory to be gauge invariant. However we have shown that if $\left[\hat{\rho}(\vec{x}),\hat{\vec{J}}(\vec{y})\right]$ is zero then the vacuum state cannot be the minimum energy state (See discussion in Section 2). Therefore the vacuum state must be defined in such a manner to allow for the existence of negative energy states with less energy than that of the vacuum. If that is not done then the theory will not be internally consistent and will not be gauge invariant.

In analyzing these problems previous researchers started out with a definition of the vacuum state $|0\rangle$ which they assumed could not be modified. When inconsistencies were found in the theory it was assumed that they were due to some mathematical difficulties associated with evaluating the commutator $\left[\hat{\rho}(\vec{x}),\hat{\vec{J}}(\vec{y})\right]$. Here we have taken a different approach. We start out with equations (1) and (56). The commutator relation given by (6) through (8) follow from (1). From these and (56) we obtain a physical theory that is gauge invariant and has local charge conservation. The next logical step is to define state vectors to be consistent with these results. That is, we need to define the state vectors so that the quantity $\langle\Omega|\left[\hat{\rho}(\vec{x}),\hat{\vec{J}}(\vec{y})\right]|\Omega\rangle$ is unambiguously zero. This requires that the vacuum state be given by $|0, \Delta E_W \to \infty\rangle$ instead of $|0\rangle$. If, instead, the state vector $|0\rangle$ is used then the gauge invariance and local charge conservation of the theory are destroyed. Of course the final result is gauge invariant and charge conserving because the unacceptable terms that appear in the perturbation expansions are identified and removed. The final result is physically correct only because ad hoc "working rules" have been applied.

In conclusion, it has been shown that Dirac field theory, in its current formulation, is mathematically inconsistent. The physical requirement of local charge conservation and gauge invariance do not follow directly from the current formulation but can only be achieved when an ad hoc set of working rules are applied. This inconsistency is a result of the notion that the vacuum state must be the state of minimum free field energy. We replace this with the requirement that the commutator $\left[\hat{\rho}(\vec{x}), \hat{\vec{J}}(\vec{y})\right]$ is unambiguously zero. This will result in a mathematically consistent theory which is gauge invariant and in which there is local charge conservation. To meet this requirement we take as the vacuum state the state vector $|0, \Delta E_W \to \infty\rangle$. When doing calculations the quantity $\Delta E_W$ is considered to be a finite number which goes to infinity as the end of the calculation.

## Appendix A

Prove

$$\left[\hat{\rho}(\vec{y}),\hat{\vec{J}}(\vec{x})\right]=0 \tag{6}$$

The above can be proven by using the following relationship from page 223 of Greiner et al [6] which holds if $\{A,C\}$, $\{A,D\}$, $\{B,C\}$, $\{B,D\}$ are c-numbers,

$$\begin{aligned}[[A,B],[C,D]] = \\ 2[A,D]\{B,C\}-2[B,D]\{A,C\}-2[A,C]\{B,D\}+2[B,C]\{A,D\}\end{aligned} \tag{A1}$$

Use (2) and (3) to obtain

$$\left[\hat{\rho}(\vec{y}),\hat{\vec{J}}(\vec{x})\right]=\left(\frac{e}{2}\right)^2\left[\left[\hat{\psi}_c^\dagger(\vec{y}),\hat{\psi}_c(\vec{y})\right],\left[\hat{\psi}_a^\dagger(\vec{x}),\vec{\alpha}_{ab}\hat{\psi}_b(\vec{x})\right]\right] \tag{A2}$$

where summation over repeated indices is implied. Use (A1) to obtain

$$\begin{aligned}\left[\left[\hat{\psi}_c^\dagger(\vec{y}),\hat{\psi}_c(\vec{y})\right],\left[\hat{\psi}_a^\dagger(\vec{x}),\vec{\alpha}_{ab}\hat{\psi}_b(\vec{x})\right]\right] &= 2\left[\hat{\psi}_c^\dagger(\vec{y}),\vec{\alpha}_{ab}\hat{\psi}_b(\vec{x})\right]\left\{\hat{\psi}_c^\dagger(\vec{y}),\hat{\psi}_a^\dagger(\vec{x})\right\} \\ &+2\left[\hat{\psi}_c(\vec{y}),\hat{\psi}_a^\dagger(\vec{x})\right]\left\{\hat{\psi}_c^\dagger(\vec{y}),\vec{\alpha}_{ab}\hat{\psi}_b(\vec{x})\right\}\end{aligned} \tag{A3}$$

where we have used the fact that the anticommutators $\left\{\hat{\psi}_c^\dagger(\vec{y}),\hat{\psi}_a^\dagger(\vec{x})\right\}$ and $\left\{\hat{\psi}_c(\vec{y}),\hat{\psi}_b(\vec{x})\right\}$ are zero. Use (1) in the above to obtain

$$\begin{aligned}\left[\left[\hat{\psi}_c^\dagger(\vec{y}),\hat{\psi}_c(\vec{y})\right],\left[\hat{\psi}_a^\dagger(\vec{x}),\vec{\alpha}_{ab}\hat{\psi}_b(\vec{x})\right]\right] &= 2\left[\hat{\psi}_c^\dagger(\vec{y}),\vec{\alpha}_{ab}\hat{\psi}_b(\vec{x})\right]\delta_{ac}\delta^3(\vec{x}-\vec{y}) \\ &+2\left[\hat{\psi}_c(\vec{y}),\hat{\psi}_a^\dagger(\vec{x})\right]\vec{\alpha}_{ab}\delta_{bc}\delta^3(\vec{x}-\vec{y})\end{aligned}$$

$$= 2\left[\hat{\psi}_a^\dagger(\vec{y}), \vec{\alpha}_{ab}\hat{\psi}_b(\vec{x})\right]\delta^3(\vec{x}-\vec{y})$$
$$+ 2\left[\vec{\alpha}_{ac}\hat{\psi}_c(\vec{y}), \hat{\psi}_a^\dagger(\vec{x})\right]\delta^3(\vec{x}-\vec{y})$$

$$= -2\left[\vec{\alpha}_{ac}\hat{\psi}_c(\vec{x}), \hat{\psi}_a^\dagger(\vec{y})\right]\delta^3(\vec{x}-\vec{y})$$
$$+ 2\left[\vec{\alpha}_{ac}\hat{\psi}_c(\vec{y}), \hat{\psi}_a^\dagger(\vec{x})\right]\delta^3(\vec{x}-\vec{y})$$

(A4)

In the last step we have used the fact that 'b' is a dummy index and replaced it by 'c'. The delta function has the property $f(t)\delta(t-a) = f(a)\delta(t-a)$ for any function f. Use this in the above expression to obtain

$$\left[\hat{\rho}(\vec{y}), \hat{\vec{J}}(\vec{x})\right] = 2\left(\frac{e}{2}\right)^2 \left(\left[\vec{\alpha}_{ac}\hat{\psi}_c(\vec{x}), \hat{\psi}_a^\dagger(\vec{x})\right]\delta^3(\vec{x}-\vec{y}) - \left[\vec{\alpha}_{ac}\hat{\psi}_c(\vec{x}), \hat{\psi}_a^\dagger(\vec{x})\right]\delta^3(\vec{x}-\vec{y})\right) = 0$$

(A5)

To prove

$$\left[\hat{\rho}(\vec{y}), \hat{\rho}(\vec{x})\right] = 0 \qquad (7)$$

replace quantities of the form $\vec{\alpha}_{ab}$ with $\delta_{ab}$ throughout the above discussion.

Next prove that

$$\left[\hat{H}_o, \hat{\rho}(\vec{x})\right] = i\vec{\nabla}\cdot\vec{J}(\vec{x}) \qquad (8)$$

From (3) and (4) we obtain

$$\left[\hat{H}_o, \hat{\rho}(\vec{x})\right] = \left(\frac{e}{4}\right)\int\left[\left[\hat{\psi}_a^\dagger(\vec{y}), H_{o,ab}\hat{\psi}_b(\vec{y})\right], \left[\hat{\psi}_c^\dagger(\vec{x}), \hat{\psi}_c(\vec{x})\right]\right]d\vec{y} \qquad (A6)$$

Use (A1) and (1) to obtain

$$\left[\left[\hat{\psi}_a^\dagger(\vec{y}), H_{o,ab}\hat{\psi}_b(\vec{y})\right],\left[\hat{\psi}_c^\dagger(\vec{x}),\hat{\psi}_c(\vec{x})\right]\right] = 2\left[\hat{\psi}_a^\dagger(\vec{y}),\hat{\psi}_c(\vec{x})\right]H_{o,ab}\delta_{bc}\delta^3(\vec{x}-\vec{y})$$
$$+2\left[H_{o,ab}\hat{\psi}_b(\vec{y}),\hat{\psi}_c^\dagger(\vec{x})\right]\delta_{ac}\delta^3(\vec{x}-\vec{y})$$

$$= -i2\left[\hat{\psi}_a^\dagger(\vec{y}),\vec{\alpha}_{ac}\hat{\psi}_c(\vec{x})\right]\vec{\nabla}\delta^3(\vec{x}-\vec{y})$$
$$-i2\left[\vec{\alpha}_{ab}\vec{\nabla}\hat{\psi}_b(\vec{y}),\hat{\psi}_a^\dagger(\vec{x})\right]\delta^3(\vec{x}-\vec{y}) \qquad \text{(A7)}$$

Integrate the above to obtain

$$2\int\begin{pmatrix} -i\left[\hat{\psi}_a^\dagger(\vec{y}),\vec{\alpha}_{ac}\hat{\psi}_c(\vec{x})\right]\cdot\vec{\nabla}\delta^3(\vec{x}-\vec{y}) \\ -i\left[\vec{\alpha}_{ab}\cdot\vec{\nabla}\hat{\psi}_b(\vec{y}),\hat{\psi}_a^\dagger(\vec{x})\right]\delta^3(\vec{x}-\vec{y})\end{pmatrix}d\vec{y} = 2\begin{pmatrix} i\left[\vec{\nabla}\hat{\psi}_a^\dagger(\vec{x}),\cdot\vec{\alpha}_{ac}\hat{\psi}_c(\vec{x})\right] \\ -i\left[\vec{\alpha}_{ab}\cdot\vec{\nabla}\hat{\psi}_b(\vec{x}),\hat{\psi}_a^\dagger(\vec{x})\right]\end{pmatrix}$$
$$= i2\vec{\nabla}\cdot\left[\hat{\psi}_a^\dagger(\vec{x}),\vec{\alpha}_{ac}\hat{\psi}_c(\vec{x})\right] \qquad \text{(A8)}$$

Use (A6) and (2) and the above result to yield (8).

<u>Appendix B</u>

Prove

$$\left[\hat{H}_o, e^{-i\hat{\rho}_w}\right] = -e^{-i\hat{\rho}_w}\int\hat{\vec{J}}\cdot\vec{\nabla}\chi \qquad (76)$$

Evaluate the following expression

$$\hat{H}_o e^{-i\hat{\rho}_w} = \hat{H}_o\sum_{n=0}^{\infty}\frac{(-i)^n(\hat{\rho}_w)^n}{n!} \qquad \text{(B1)}$$

From (12) and (10)

$$\hat{H}_o(\hat{\rho}_w)^n = (\hat{H}_o\hat{\rho}_w)(\hat{\rho}_w)^{n-1} = \left(\hat{\rho}_w H_o - i\int\hat{\vec{J}}\cdot\vec{\nabla}\chi\, d\vec{x}\right)(\hat{\rho}_w)^{n-1}$$
$$= \hat{\rho}_w H_o(\hat{\rho}_w)^{n-1} - i(\hat{\rho}_w)^{n-1}\int\hat{\vec{J}}\cdot\vec{\nabla}\chi\, d\vec{x}$$
$$\text{(B2)}$$

Repeatedly applying (12) and (10) to move $H_o$ to the right yields

$$\hat{H}_o(\hat{\rho}_w)^n = H_o(\hat{\rho}_w)^n - in(\hat{\rho}_w)^{n-1}\int \hat{\vec{J}}\cdot\vec{\nabla}\chi\, d\vec{x} \tag{B3}$$

Use this in (B1) to obtain

$$\hat{H}_o e^{-i\hat{\rho}_w} = \sum_{n=0}^{\infty}\frac{(-i)^n}{n!}\left((\hat{\rho}_w)^n\hat{H}_o - in(\hat{\rho}_w)^{n-1}\int \hat{\vec{J}}\cdot\vec{\nabla}\chi\, d\vec{x}\right) = e^{-i\hat{\rho}_w}\left(\hat{H}_o - \int \hat{\vec{J}}\cdot\vec{\nabla}\chi\, d\vec{x}\right) \tag{B4}$$

which is the same as (76).